\documentclass[12pt]{article}
\usepackage{amsmath}
\usepackage{bm}
\usepackage{color}
\usepackage{braket}
\usepackage{caption}
\usepackage{subcaption}
\usepackage{graphicx}

\begin{document}
\title{Detection of energy levels of a spin system on a quantum computer by probe spin evolution}

\author{
Kh. P. Gnatenko$^{1,2}$\footnote{khrystyna.gnatenko@gmail.com}, H. P. Laba$^3$\footnote{hanna.p.laba@lpnu.ua},
V. M. Tkachuk$^1$\footnote{voltkachuk@gmail.com}\\
$^1$ Professor Ivan Vakarchuk Department for Theoretical Physics,\\
Ivan Franko National University of Lviv,\\
12, Drahomanov St., Lviv, 79005, Ukraine.\\
$^2$ SoftServe Inc., 2d Sadova St., Lviv, 79021, Ukraine.\\
$^3$Department of Applied Physics and Nanomaterials Science, \\
Lviv Polytechnic National University,\\
5 Ustiyanovych St., 79013 Lviv, Ukraine.}

\maketitle

\begin{abstract}
We propose a method for detection of energy levels of arbitrary spin system on a quantum computer based on studies of evolution of only one probe spin. On the basis of the proposed method  energy levels of spin systems are found on IBM's quantum computer ibmq-bogota, among them are spin chain in magnetic field, triangle spin cluster, Ising model on squared lattice in magnetic field. The results of quantum calculations are in agreement with the theoretical ones.  The method is efficient for estimation of the energy levels of many-spin systems and opens a possibility to achieve quantum supremacy in solving eigenvalue problem with development of multi-qubit quantum computers.
\end{abstract}

\section{Introduction}
Studying of energy levels of  physical systems is one of the central problems of quantum mechanics that can be solved on a quantum computer.
To estimate the  energies of a quantum system the quantum phase estimation algorithm was developed  \cite{Kitaev95,Abrams97,Abrams99,Kitaev97}. The algorithm has been widely used (see, for instance, \cite{Abrams97,Abrams99,Kitaev97,Dob,Paesani,Parker20, Cruz20, Russo} and references therein). For detecting of energies of transitions the robust phase estimation algorithm was introduced in \cite{Russo}.

Method for estimation of energy levels by time dependence of expectation values of the evolution operator
was proposed \cite{Somma19,Somma02}. The method is based on quantum algorithm with controlled operator of evolution (see \cite{Somma19}). 
The spectroscopy
protocol  to extract many-body spectra in
experimental simulations was introduced in \cite{Bryce}. The protocol diabatically ramps the transverse magnetic field to create excitations. 
In \cite{Motta}  quantum Lanczos and quantum imaginary time evolution algorithms were proposed. 
 Qubit efficient scheme to study ground-state properties of quantum many-body systems was suggested in \cite{Liu}.
Well known method allowing to find  transition energies is  classical-quantum algorithm variational quantum eigeinsolver \cite{Peruzzo,Mc,Malley,Parrish}. Also the ground state and therefore the energy of the  state can be found with  quantum approximate optimization algorithm  \cite{Farhi, Farhi1,Moll,Fuchs}.

In our recent paper \cite{Gna2021} we proposed a method for detecting of the energy levels of a quantum system on  a quantum computer based on studies of evolution of mean value of operator anticommuting with Hamiltonian of the system. It is worth noting that existence of operator anticommuting with Hamiltonian is the evidence of symmetry of its energy levels with respect to  $E\to-E$. The method was realized for detection of the energy levels of  spin systems with energy levels symmetric  with respect to  $E\to-E$  (spin in magnetic field, spin chain, Ising model on squared lattice) on IBM's quantum computers \cite{Gna2021}.

In the present paper we propose an efficient method for detecting of energy levels of arbitrary spin systems  on a quantum computer. The method is based on studies of evolution of probe spin. We realized the method for examining of the energy levels of a spin chain in the magnetic field, triangle spin cluster, Ising model on squared lattice in the magnetic field.  The results of calculations of the energy levels of the spin systems on IBM's quantum computer correspond to analytical ones.

The paper is organized as follows. In Section 2 the method for detecting of the energy levels of a quantum system on the basis of studies of evolution of probe spin is presented. Section 3 is devoted to  studies of energy levels of spin systems described by the Ising model on the basis of the proposed method. Results of quantum calculations of the energy levels of spin systems  on ibmq-bogota are presented in Section 4.  Conclusions are done in Section 5.

\section{Evolution of mean value of a probe spin and energy levels of a spin system}

We study eigenvalue problem for Hamiltonian $H$
\begin{eqnarray} \label{EigH}
H|\psi\rangle=E|\psi\rangle.
\end{eqnarray}
Energy levels are restricted from the bottom. Therefore we can shift them to the positive ones by adding  constant $C$ to the Hamiltonian.

Let us add to system under consideration additional spin (ancila qubit) and construct total Hamiltonian in the following form
\begin{eqnarray}
H_T=\sigma_0^z (H+C).\label{total}
\end{eqnarray}
Note that $[\sigma_0^z,H]=0$. Eigenvalues of operator $\sigma_0^z$ are $\pm 1$.
Therefore the  energy spectrum of the total Hamiltonian $H_T$ contains
positive eigenvalues of $H+C$ and eigenvalues of $-(H+C)$ that are negative. Note that the energy spectrum of $H_T$ is symmetric with respect to $E_T\to-E_T$ (here $E_T$ are the energy levels of $H_T$). It is easy to write operators  anticommuting with the total Hamiltonian. We have
\begin{eqnarray}
\{\sigma_0^x, H_T\}=\{\sigma_0^y, H_T\}=0.
\end{eqnarray}
Let us consider mean value of $\sigma_0^x$ in evolution governed by $H_T$.
Starting from  initial state $|\psi_0\rangle$ we find
\begin{eqnarray}
\langle\sigma_0^x(t)\rangle=\langle\psi_0|e^{iH_Tt/\hbar} \sigma_0^x e^{-iH_Tt/\hbar}|\psi_0\rangle=\nonumber\\
=\langle\psi_0|e^{2iH_Tt/\hbar} \sigma_0^x |\psi_0\rangle=
\langle\psi_0|\sigma_0^x e^{-2iH_Tt/\hbar}  |\psi_0\rangle. \label{EEE}
\end{eqnarray}
To write (\ref{EEE}) we take into account that $\sigma_0^x e^{-iHt/\hbar}= e^{iHt/\hbar}\sigma_0^x$.
We expand the initial state over the eigeinstates  of the  total Hamiltonian $|E^i_T\rangle$ corresponding to the  energies $E^i_T$, namely, $|\psi_0\rangle=\sum_{i}c_i|E^i_T\rangle$. Therefore for $\langle\sigma_0^x(t)\rangle$ we  write
\begin{eqnarray}
\langle\sigma_0^x(t)\rangle=\sum_j g_je^{-i2\omega^j_Tt},
\end{eqnarray}
with $g_j$ given by
$g_j=\sum_ic_i^*\langle E^i_T|\sigma_0^x|E^j_T\rangle c_j$ and $\omega^i_T=E^i_T/\hbar$. 

To detect the energy levels of the system $H$ we consider Fourier transformation for $\langle\sigma_0^x(t)\rangle$. We obtain
\begin{eqnarray}\label{Aomega}
\sigma^x_0(\omega)={1\over 2\pi}\int_{-\infty}^{\infty}dt \langle\sigma_0^x(t)\rangle e^{i\omega t}=
\sum_i g_i\delta(\omega-2\omega^{i}_T).
\end{eqnarray}
It is important to stress that the function $\sigma^x_0(\omega)$ has $\delta$ - peaks at  $\omega=2\omega^{i}_T$. corresponding to the energy levels $E^i_T=\hbar\omega^i_T$.

 On a quantum computer we can detect values of $\langle\sigma_0^x(t)\rangle$ at fixed moments of time.
Considering $t=\tau n$, were $n=-N, -N+1,...N-1, N$ with  $\tau$ being time interval, similarly as in \cite{Gna2021} we find
\begin{eqnarray} \nonumber
\sigma_0^x(\omega)={\tau\over 2\pi}\sum_jg_j\sum_{n=-N}^N  e^{-i(\omega-2\omega^{j}_T)\tau n}=\\
={\tau\over 2\pi}\sum_jg_j\left( 1 + 2\cos((\omega-2\omega^{j}_T)(N+1)\tau/2)  {\sin((\omega-2\omega_T^{j})N\tau/2)\over\sin((\omega-2\omega_T^{j})\tau/2)}\right).
\end{eqnarray}
 In the limit $\tau\to 0$ and fixed $N\tau$ we can write
\begin{eqnarray} \label{AomegaT}
\sigma_0^x(\omega)={1\over\pi}\sum_jg_j{\sin((\omega-2\omega^{j}_T)N\tau)\over(\omega-2\omega^{j}_T)}.
\end{eqnarray}
Function (\ref{AomegaT}) has peaks at $\omega=2\omega^{j}_T$. Therefore the energy levels of the total Hamiltonian $H_T$ and the energy levels of the system with Hamiltonian $H$ can be detected on the basis of studies of $\sigma_0^x(\omega)$. Taking into account (\ref{total}) the energy levels of  $H$ can be found as $E=E_T-C$, where $E_T$ are positive energies of $H_T$.

To detect  all peaks corresponding to the
eigeinvalues of $H_T$  in the Fourier transformation  the time interval $\tau$ has to be chosen to satisfy the following inequality  $\tau = \pi/2\omega^{max}_T$. Here $\omega^{max}_T$ correspond to the maximal eigeinvalue of $H_T$  $E_T^{max}=\hbar\omega^{max}_T$ \cite{Nyquist}.
The delta-peaks are more thin and more higher for larger $N\tau$.

\section{Detecting energy levels of Ising model on a quantum computer}

We consider $N$ spins described by the Ising model. The Hamiltonian reads
\begin{eqnarray}
H={1\over 2}\sum_{i,j}J_{ij}\sigma^z_i\sigma^z_j,\label{Ising}
\end{eqnarray}
where  $J_{ij}$ represents interaction between spins.

The total system with additional spin is represented by the following Hamiltonian
\begin{eqnarray}
H_T=\sigma_0^z (H+C)={1\over 2}\sum_{i,j}J_{ij}\sigma_0^z\sigma^z_i\sigma^z_j+C\sigma_0^z,
\end{eqnarray}
where constant $C$ is added to shift the energy spectrum of the Hamiltonian of a system $H$ to the positive values.

Let us consider evolution of state vector governed by $H_T$. As an initial state we choose
\begin{eqnarray}\label{psi0}
|\psi_0\rangle=|++...+\rangle=H^{[N+1]}|00...0\rangle=
{1\over 2^{\frac{N+1}{2}}}\sum_{x_0x_1x_2...x_N}|x_0x_1x_2...x_N\rangle,
\end{eqnarray}
where $H^{[N+1]}=\prod_{i=0}^N H_i $, $H_i$ is the Hadamard operator acting on $i$-th qubit, $x_i=0,1$.
Qubit is associated with spin, namely, $|0\rangle=|\uparrow\rangle$ and
$|0\rangle=|\downarrow\rangle$.

The evolution of the mean value of $\sigma_0^x$ reads
\begin{eqnarray}\label{At}
\langle\sigma_0^x(t)\rangle =\langle + ++...+|
\sigma_0^x e^{-i2H_Tt/\hbar}|+++...+\rangle
={1\over 2^{N+1}}\times\nonumber\\\times\sum_{x_0,x_1,x_2,...x_N}\langle x_0x_1x_2...x_N|
e^{-i2H_Tt/\hbar}|x_0x_1x_2...x_N\rangle
={1\over 2^{N+1}}\sum_{k=1}^{2^{N+1}}e^{-i2\omega^k_Tt}.
\end{eqnarray}
Here we
take into account  that $\sigma_0^x|+\rangle=|+\rangle$, $|x_0x_1x_2...x_N\rangle$
are the eigenstates of the total hamiltonian $H_T$, and
$\omega^k_T$ are the energy levels in the unit of $\hbar$.
To extract the frequencies
$\omega_T^{i}$ we consider the Fourier transformation (\ref{Aomega}) 
\begin{eqnarray}\label{Adelta}
\sigma_0^x(\omega)={1\over 2^{N+1}}\sum_i\delta(\omega-2\omega^{i}_T),
\end{eqnarray}
here $g_i={1/2^{N+1}}$.
Thus function $\sigma(\omega)$ has $\delta$ - peaks at $\omega=2\omega^{i}_T$. It allows
to find the frequencies $\omega^{i}_T$ that correspond to energy levels $E_i=2\hbar\omega^i_T$ of the total Hamiltonian.
Positive part of this spectrum corresponds to the energy levels of the Hamiltonian of the system $H$ (\ref{Ising}) shifted on the constant $C$.

\begin{figure}[!!h]
\begin{center}
\includegraphics[scale=0.55, angle=0.0, clip]{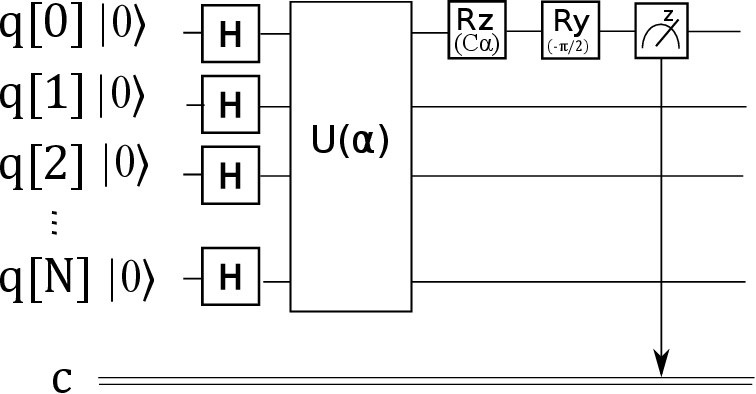}
\end{center}
\caption{Quantum protocol for studies of evolution of mean value of a probe spin on a quantum computer, $U(\alpha)=\exp{(-i\alpha \sigma_0^zH/2)}$, $\alpha=2t/\hbar$.}
		\label{fig:7}
\end{figure}

Quantum protocol for studies of evolution of mean value of a probe spin is presented on Fig. \ref{fig:7}.
In the protocol  $\alpha=2t/\hbar$, $U(\alpha)=\exp{(-i\alpha \sigma_0^zH/2)}$. To quantify the mean value of $\sigma^x_0$ we take into account that the operator $\sigma^x_0$ can be represented as   $\sigma^x_0=\exp(-i\pi\sigma^y_0/4)\sigma^z_0\exp(i\pi\sigma^y_0/4)$. Therefore before measurement in the standard basis the state of the corresponding qubit has to be rotated with  $RY(-\pi/2)$ gate.

In the next section using the proposed quantum protocol Fig. \ref{fig:7} we study the energy levels of spin systems on IBM's quantum computer.

\section{Estimation of the energy levels of spin systems by studies of a probe spin evolution on IBM's quantum computer}

\subsection{Spin chain in  magnetic field}

 Let us consider a spin chain in magnetic field described by the following Hamiltonian
 \begin{eqnarray}
H=J\sigma^{z}_1\sigma^{z}_2+J\sigma^{z}_2\sigma^{z}_3+J\sigma^{z}_1+J\sigma^{z}_2+J\sigma^{z}_3.\label{hhh}
\end{eqnarray}
In this case the total Hamiltonian (\ref{total}) reads
\begin{eqnarray}\label{Adelta00}
H_T=J\sigma^{z}_0(\sigma^{z}_1\sigma^{z}_2+\sigma^{z}_2\sigma^{z}_3+\sigma^{z}_1+\sigma^{z}_2+\sigma^{z}_3+C).
\end{eqnarray}
To provide the positivity of the energy levels of $H$ the constant $C$ is chosen to be $C=6J$. The initial state is considered as  $\ket{+++}$.
We detect the mean value of $\sigma^x_0$ on IBM's quantum computer  ibmq-bogota.
For convenience we put $J/\hbar=1$ and perform calculations for $\alpha/2=Jt/\hbar$  changing from $-8\pi$ to $8\pi$ with the step $\pi/24$. 

\begin{figure}[!!h]
\begin{center}
\includegraphics[scale=0.55, angle=0.0, clip]{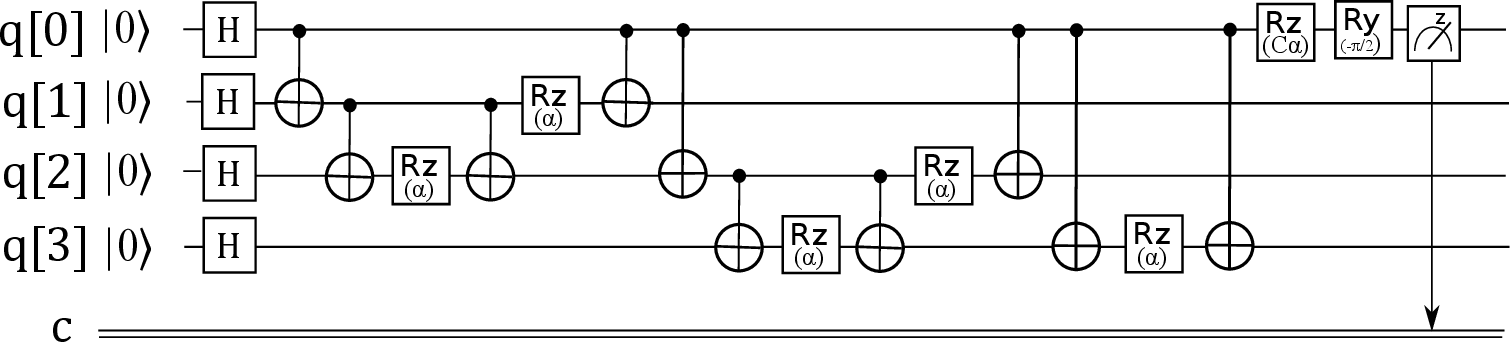}
\end{center}
\caption{Quantum protocol for studies of evolution of mean value of $\sigma^x_0$ in the case of spin chain in the magnetic field (\ref{hhh}) on a quantum computer. Here  $C=6J$ and $\alpha=2Jt/\hbar$. }
		\label{prot_chain}
\end{figure}

The quantum protocol for the studies is given on Fig. \ref{prot_chain}.    Constructing the quantum protocol we take into account that with exactness to total phase factor operator $\exp(-iJ\sigma_j^z\sigma_k^zt/\hbar)$ can be represented as $CNOT_{jk}RZ_k(2Jt/\hbar)CNOT_{jk}$, here $RZ_k(2Jt/\hbar)$ is the Z-rotation gate acting on  qubit $q[k]$, $CNOT_{jk}$ is the controlled-NOT gate (operator $CNOT_{jk}$ acts on $q[j]$ as control and on $q[k]$ as a target). Also,  with exactness to total phase factor we use representation of $\exp(-iJ\sigma_j^z\sigma_k^z\sigma_l^zt/\hbar)$  as $CNOT_{jk}CNOT_{kl}RZ_l(2Jt/\hbar)CNOT_{kl}CNOT_{jk}$ and take into account that $(CNOT_{jk})^2=1$.

\begin{figure}
\begin{center}
\subcaptionbox{\label{ff2}}{\includegraphics[scale=0.75, angle=0.0, clip]{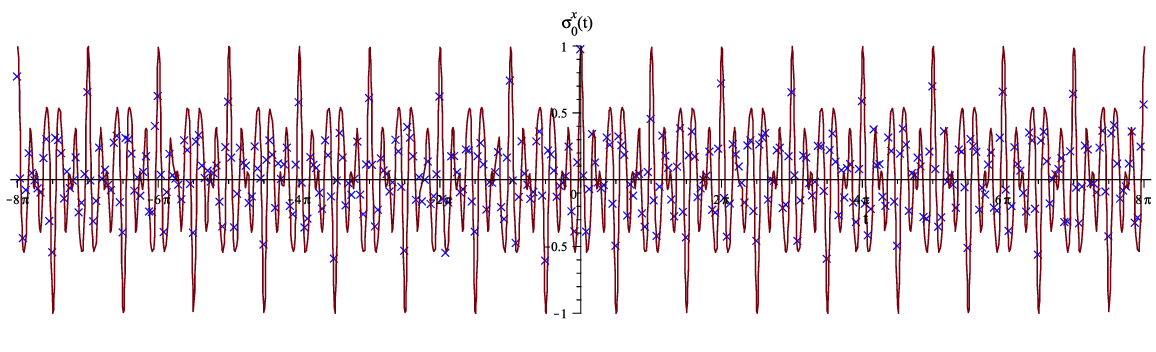}}
\subcaptionbox{\label{ff2}}{\includegraphics[scale=0.3, angle=0.0, clip]{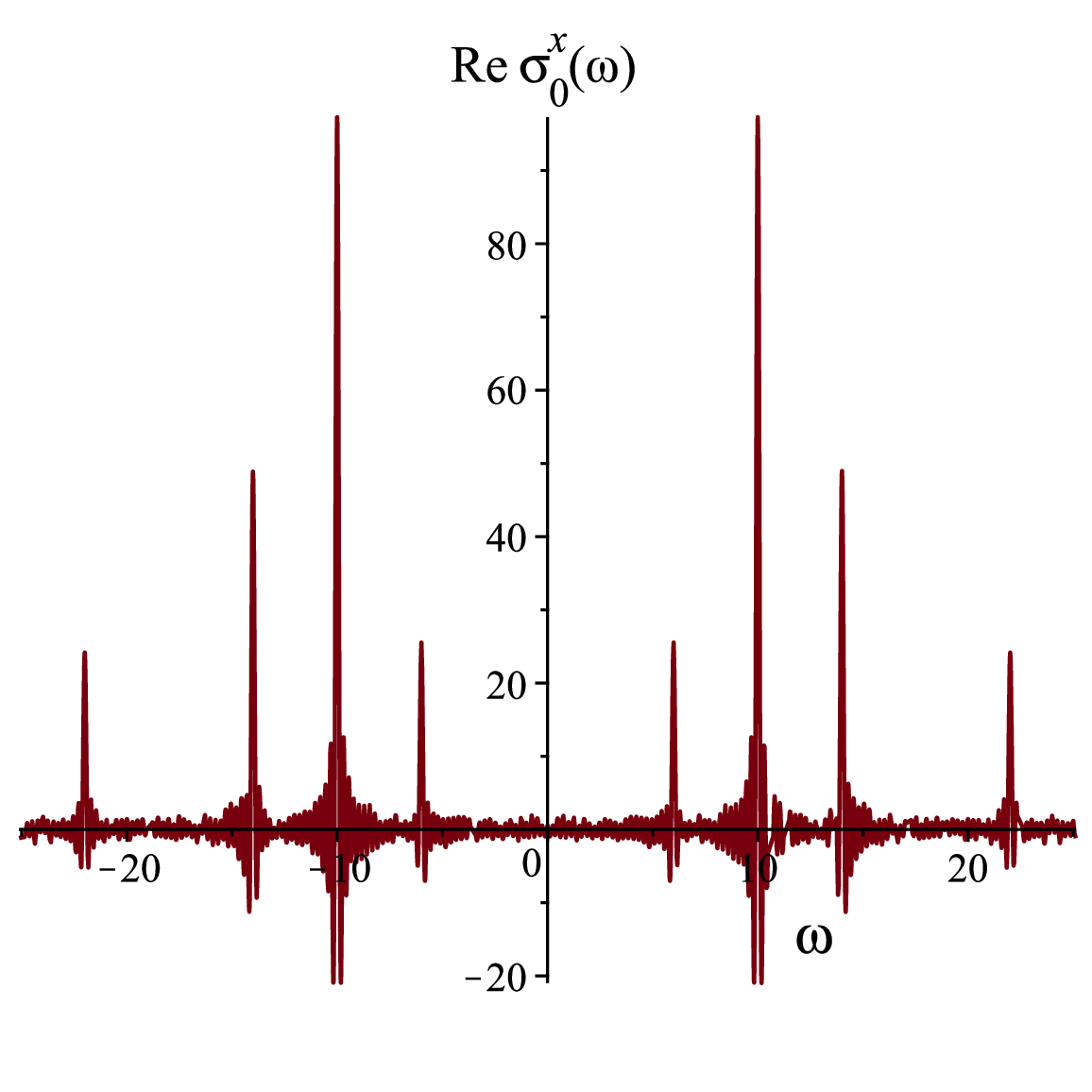}}
\subcaptionbox{\label{ff2}}{\includegraphics[scale=0.3, angle=0.0, clip]{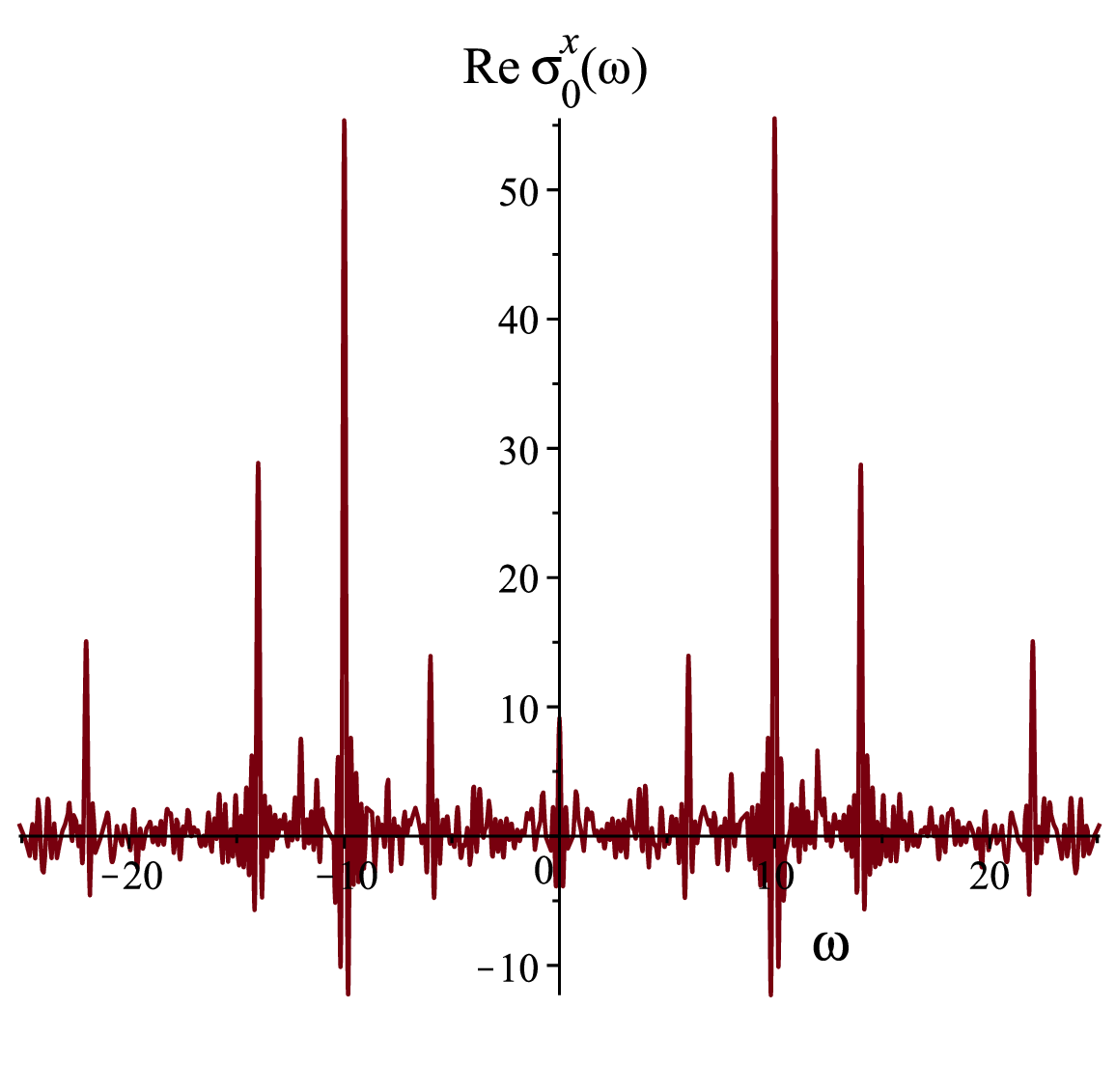}}
\end{center}
\caption{Results of quantifying  of the energy levels of  spin chain  in the magnetic field (\ref{hhh}). Evolution of the mean value $\braket{\sigma^x_0}$ detected on ibmq-bogota (a).   The real part of $\sigma^x_0(\omega)$ obtained on the basis of the results of calculation of $\sigma^x_0(t)$ on ibmq-qasm-simulator (b)  and  on ibmq-bogota (c).   The peaks of $\textrm{Re} \,\sigma^x_0(\omega)$  at $\omega=\pm6$, $\omega=\pm 10$, $\omega=\pm 14$, $\omega=\pm 22$  correspond to energies $E_T=\pm3J$, $E_T=\pm5J$, $E=\pm 7 J$, $E=\pm 11J$ of the total Hamiltonian (\ref{Adelta00}) and  energies $E=-3J$, $E=-J$, $E=J$, $E=5J$ of the spin chain in the magnetic field (\ref{hhh}).}
		\label{fig99}
\end{figure}

The results of detecting of the energy levels of the spin chain in the magnetic field are presented on Fig. \ref{fig99}.
The value $\sigma^x_0(\omega)$ is real (see (\ref{AomegaT})). Because of errors of quantum calculations we obtain imaginary part of $\sigma^x_0(\omega)$ which looks as a noise. Therefore on Fig. \ref{fig99} we present the real part of  $\sigma^x_0(\omega)$ detected on the quantum device. The sharp peaks of  $\textrm{Re} \,\sigma^x_0(\omega)$  at points $\omega=\pm6$, $\omega=\pm 10$, $\omega=\pm 14$, $\omega=\pm 22$ correspond to the energies $E_T=\pm3J$, $E_T=\pm5J$, $E_T=\pm7J$, $E_T=\pm11J$  of the total Hamiltonian $H_T$ (\ref{Adelta00}).

Small peak at $\omega=0$  appears in $\textrm{Re} \,\sigma^x_0(\omega)$ obtained on the basis of quantum calculations on  ibmq-bogota  Fig. \ref{fig99} (c) and is absent on  Fig. \ref{fig99} (b) corresponding to results obtained on the quantum simulator. This peak is related with systematic error shift of the results of quantum calculations for  $\sigma^x_0(t)$ Fig. \ref{fig99} (a).  Energy spectrum of the total Hamiltonian  (\ref{Adelta00})  contains
positive eigenvalues of $H+C$ and eigenvalues of $-(H+C)$ which are negative and  does not contain $E_T=0$. So, the peak at $\omega=0$  has not to be taken into account.
Using relation of $H_T$ (\ref{Adelta00}) with the Hamiltonian of the system $H$(\ref{hhh}) we find the following results for the energy levels of the spin chain in the magnetic field   $E=-3J$, $E=-J$, $E=J$, $E=5J$. The obtained spectrum corresponds to the analytical one.

 \subsection{Triangle spin cluster}

\begin{figure}[!!h]
\begin{center}
\includegraphics[scale=0.45, angle=0.0, clip]{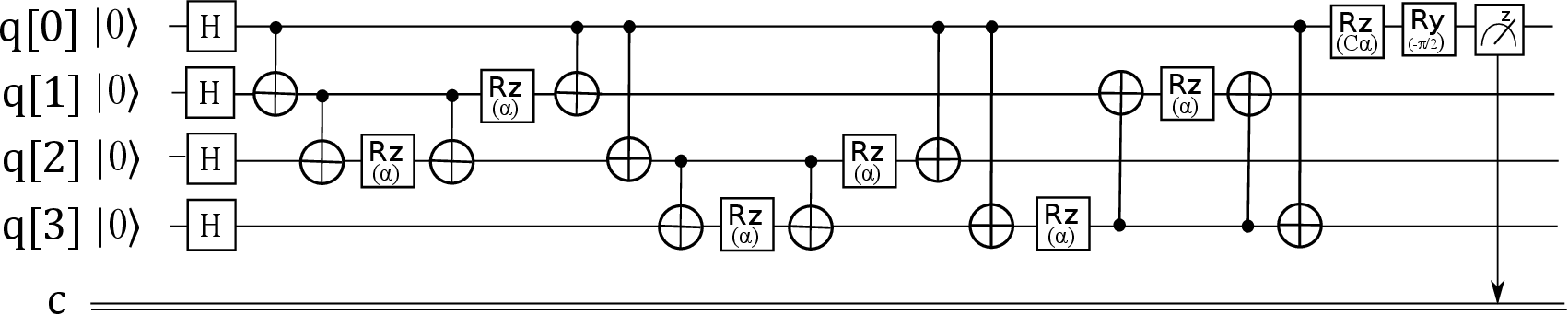}
\end{center}
\caption{Quantum protocol for studies of evolution of mean value of $\sigma^x_0$ in the case of triangle spin cluster (\ref{hhh3}) on a quantum computer. Here $C=7J$, $\alpha=2Jt/\hbar$. }
		\label{fig990}
\end{figure}

As the next example we consider triangle spin cluster described by the following Hamiltonian
 \begin{eqnarray}
H=J\sigma^{z}_1\sigma^{z}_2+J\sigma^{z}_2\sigma^{z}_3+J\sigma^{z}_1\sigma^{z}_3+J\sigma^{z}_1+J\sigma^{z}_2+J\sigma^{z}_3.\label{hhh3}
\end{eqnarray}
Choosing  $C=7J$, we can write the expression for the total Hamiltonian as 
\begin{eqnarray}\label{Adelta}
H_T=J\sigma^{z}_0(\sigma^{z}_1\sigma^{z}_2+\sigma^{z}_2\sigma^{z}_3+\sigma^{z}_1\sigma^{z}_3+\sigma^{z}_1+\sigma^{z}_2+\sigma^{z}_3+7).
\end{eqnarray}
 Quantum protocol for studies of the mean value of $\sigma^x_0$ in the case of the triangle spin cluster  is presented on Fig \ref{fig990}.

\begin{figure}
\begin{center}
\subcaptionbox{\label{ff2}}{\includegraphics[scale=0.75, angle=0.0, clip]{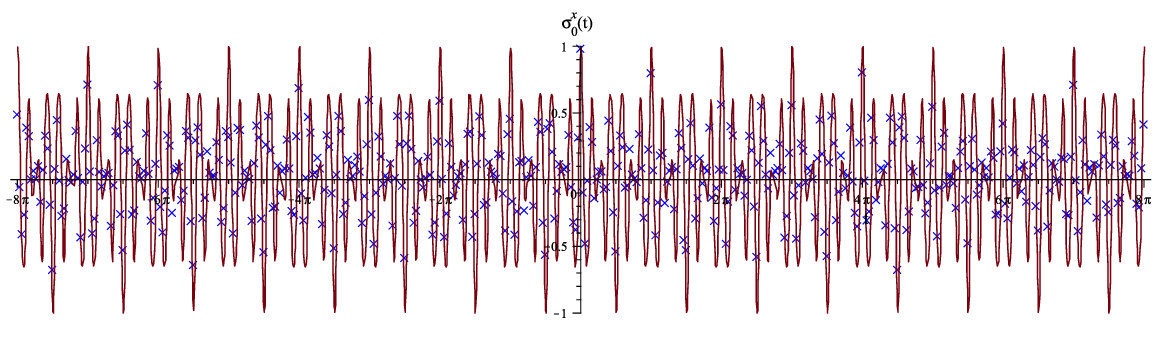}}
\subcaptionbox{\label{ff2}}{\includegraphics[scale=0.3, angle=0.0, clip]{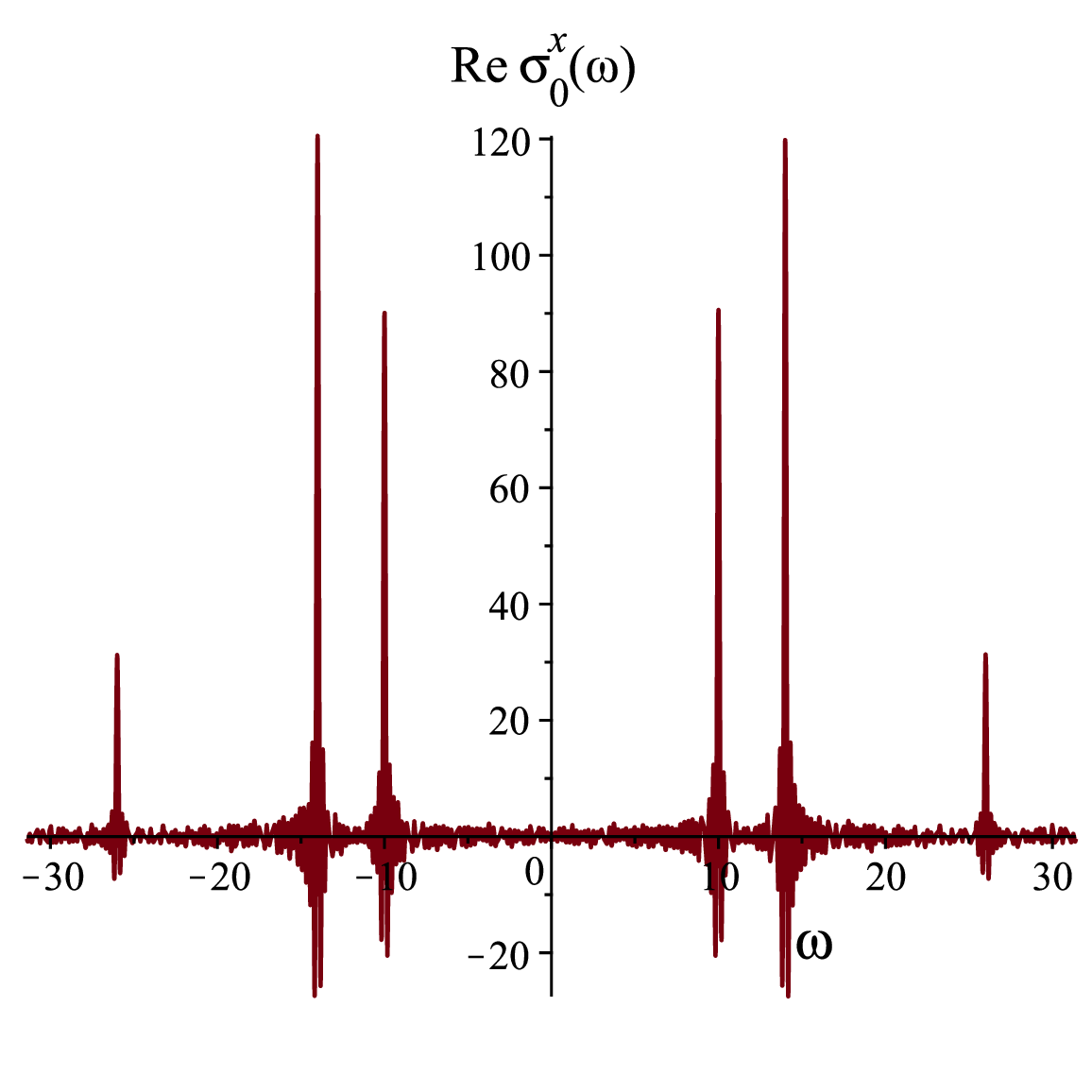}}
\subcaptionbox{\label{ff2}}{\includegraphics[scale=0.35, angle=0.0, clip]{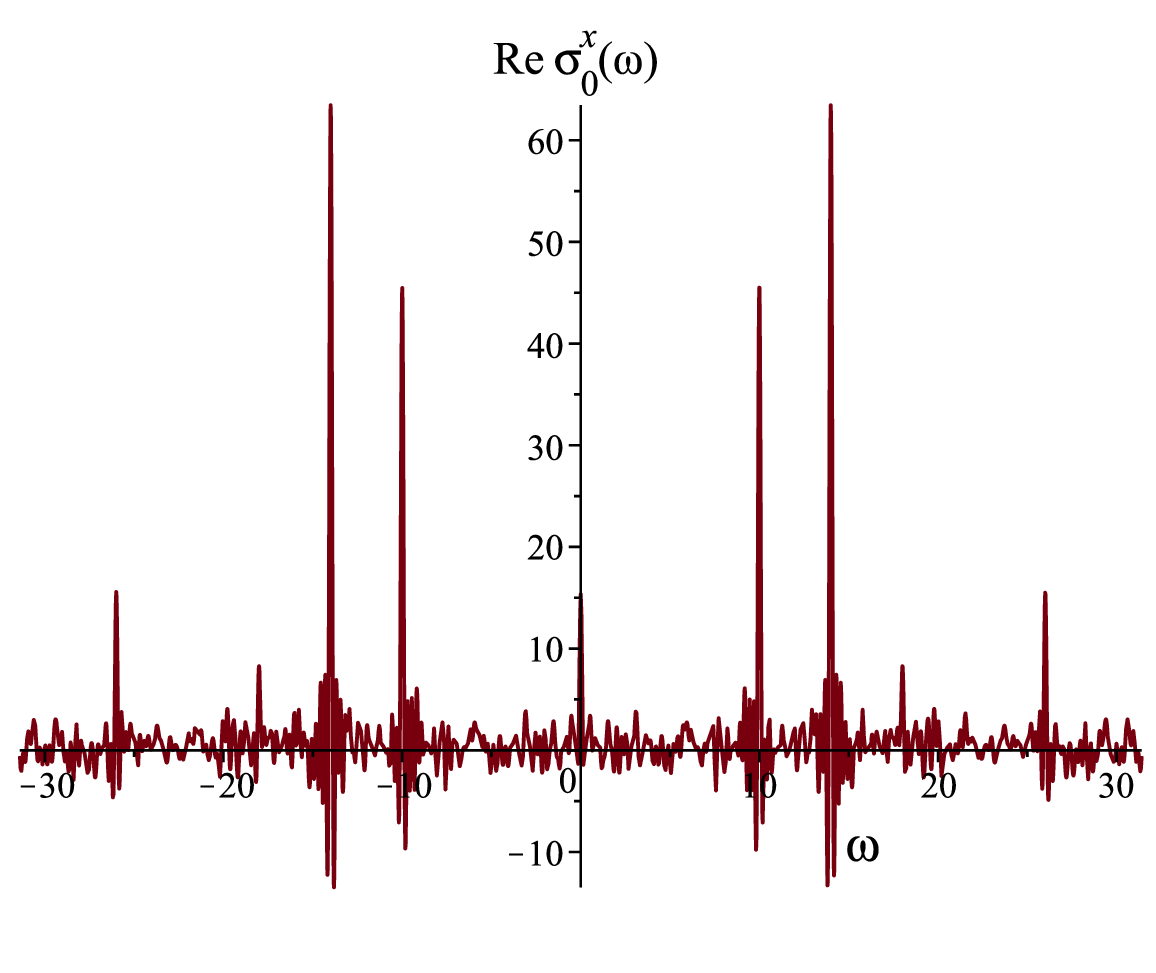}}
\end{center}
\caption{Results of quantifying  of the energy levels of  triangle spin cluster (\ref{hhh3}). Evolution of the mean value $\braket{\sigma^x_0}$ detected on ibmq-bogota  (a). The real part of $\sigma^x_0(\omega)$ obtained on the basis of calculations of $\sigma^x_0(t)$ on ibmq-qasm-simulator (b)  and  on ibmq-bogota (c).   The peaks of $\textrm{Re} \,\sigma^x_0(\omega)$ at $\omega=\pm10$, $\omega=\pm 14$, $\omega=\pm 26$ correspond to energies $E_T=\pm5J$, $E_T=\pm7J$, $E_T=\pm13J$ of the total Hamiltonian (\ref{Adelta}) and energies   $E=-2J$, $E=0$, $E=6J$ of the triangle spin cluster (\ref{hhh3}).}
		\label{fig1099}
\end{figure}

The results of quantum calculation of the mean value of $\sigma^x_0$ for $\alpha/2=Jt/\hbar$  changing from $-8\pi$ to $8\pi$ with the step $\pi/30$ are presented on Fig. \ref{fig1099} (a).
 On Fig. \ref{fig1099}  we present   $\textrm{Re} \,\sigma^x_0(\omega)$ obtained on the basis of calculation on  ibmq-qasm-simulator (b) and on ibmq-bogota (c).  The sharp peaks of  $\textrm{Re} \,\sigma^x_0(\omega)$ at the points $\omega=\pm10$, $\omega=\pm 14$, $\omega=\pm 26$, correspond to the total  Hamiltonian $H_T$ energies $E_T=\pm5J$, $E_T=\pm7J$, $E_T=\pm13J$, respectively. Therefore  for the triangle spin cluster  we obtain energy levels as follows  $E=-2J$, $E=0$, $E=6J$. The result is in agreement with the theoretical one.  Similarly as in the previous example, the peak at $\omega=0$ Fig. \ref{fig1099} (c) is related with quantum errors and has not to be taken into account.

\begin{figure}
\begin{center}
\subcaptionbox{\label{ff2}}{\includegraphics[scale=0.75, angle=0.0, clip]{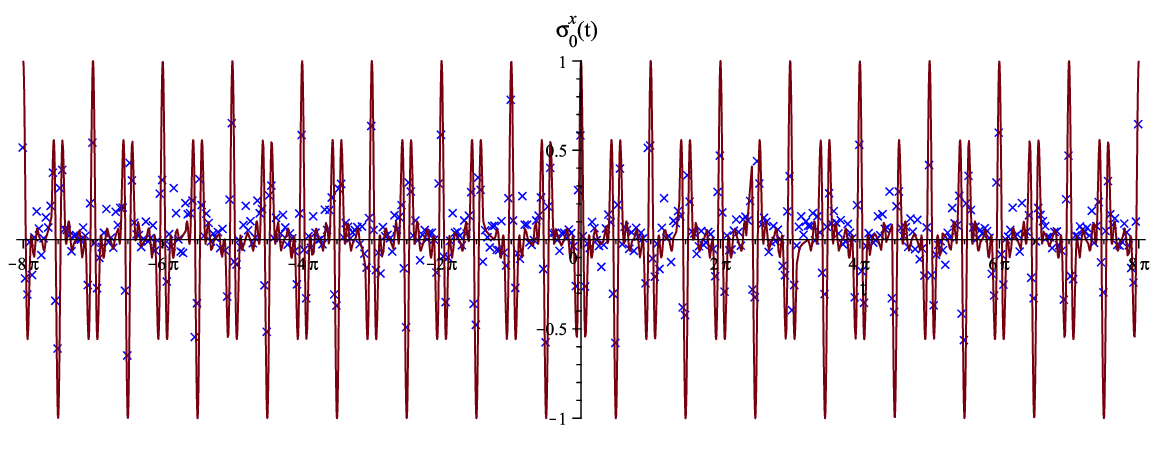}}
\subcaptionbox{\label{ff2}}{\includegraphics[scale=0.3, angle=0.0, clip]{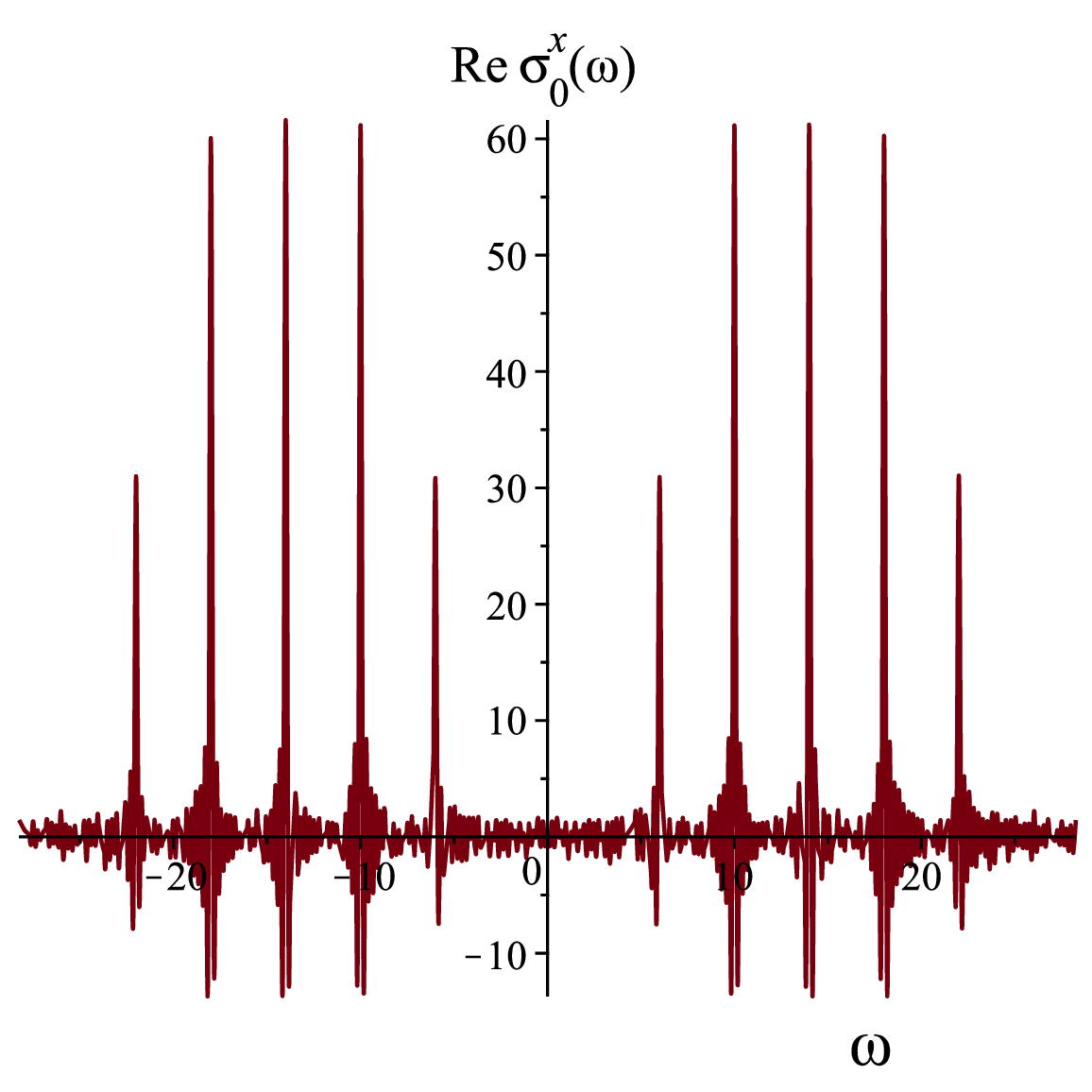}}
\subcaptionbox{\label{ff2}}{\includegraphics[scale=0.3, angle=0.0, clip]{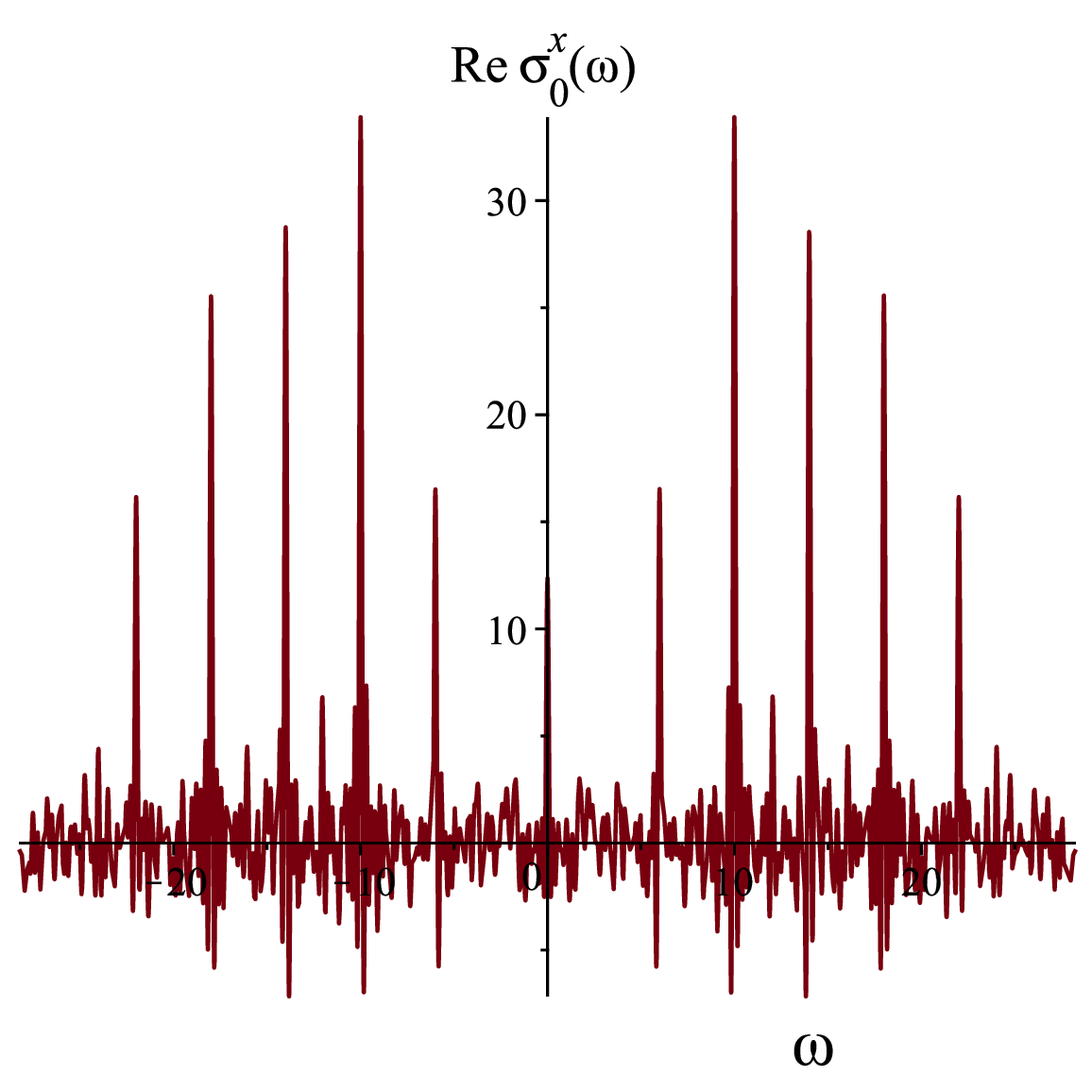}}
\end{center}
\caption{Results of quantifying  of the energy levels of  triangle spin cluster with spatial anisotropic interaction (\ref{hhh33}). Evolution of the mean value $\braket{\sigma^x_0}$ detected on ibmq-bogota (a).  The real part of $\sigma^x_0(\omega)$ obtained on the basis of calculations of $\sigma^x_0(t)$  on ibmq-qasm-simulator (b)  and  on ibmq-bogota (c).   The peaks of $\textrm{Re} \,\sigma^x_0(\omega)$ at $\omega=\pm6$, $\omega=\pm 10$, $\omega=\pm 14$, $\omega=\pm 18$,  $\omega=\pm 22$  correspond to energies $E_T=\pm3J$, $E_T=\pm5J$, $E_T=\pm7J$, $E_T=\pm11J$  of the total Hamiltonian (\ref{333Adelta}) and energies   $E=-4J$, $E=-2J$, $E=0$, $E=4J$ of the triangle spin cluster with spatial anisotropic interaction (\ref{hhh33}).}
		\label{fig0099}
\end{figure}

In addition we consider triangle spin cluster with spatial anisotropic interaction described by the  Hamiltonian
\begin{eqnarray}
H=-J\sigma^{z}_1\sigma^{z}_2+J\sigma^{z}_2\sigma^{z}_3+J\sigma^{z}_1\sigma^{z}_3+J\sigma^{z}_1+J\sigma^{z}_2+J\sigma^{z}_3.\label{hhh33}
\end{eqnarray}
Similarly as in the previous case, choosing  $C=7J$ the expression for the total Hamiltonian  reads
\begin{eqnarray}\label{333Adelta}
H_T=J\sigma^{z}_0(-\sigma^{z}_1\sigma^{z}_2+\sigma^{z}_2\sigma^{z}_3+\sigma^{z}_1\sigma^{z}_3+\sigma^{z}_1+\sigma^{z}_2+\sigma^{z}_3+7).
\end{eqnarray}
The evolution of the mean value of $\sigma^x_0$ was studied  for $\alpha/2=Jt/\hbar$  changing from $-8\pi$ to $8\pi$ with the step $\pi/30$ using quantum protocol Fig. \ref{fig990} with changing the sign  of the parameter $\alpha$ in the first $RY$ gate acting on $q[2]$ to the opposite one.  The obtained results are presented on Fig. \ref{fig0099}. On Fig.  \ref{fig0099} (b), (c) we see
peaks of $\textrm{Re} \,\sigma^x_0(\omega)$  at $\omega=\pm6$, $\omega=\pm 10$, $\omega=\pm 14$, $\omega=\pm 18$,  $\omega=\pm 22$. They correspond to  energies $E_T=\pm3J$, $E_T=\pm5J$, $E_T=\pm7J$, $E_T=\pm11J$  of the total Hamiltonian (\ref{333Adelta}), respectively.   So, on the basis of the quantum calculations we obtain the energy levels of the triangle spin cluster with spatial anisotropic interaction (\ref{hhh33}) as follows $E=-4J$, $E=-2J$, $E=0$, $E=4J$ as it should be according to the analytical calculations. The peak at $\omega=0$ Fig. \ref{fig0099}  (c) is related with quantum errors and has not to be considered.

 \subsection{Ising model on squared lattice in the magnetic field}

\begin{figure}[!!h]
\begin{center}
\includegraphics[scale=0.35, angle=0.0, clip]{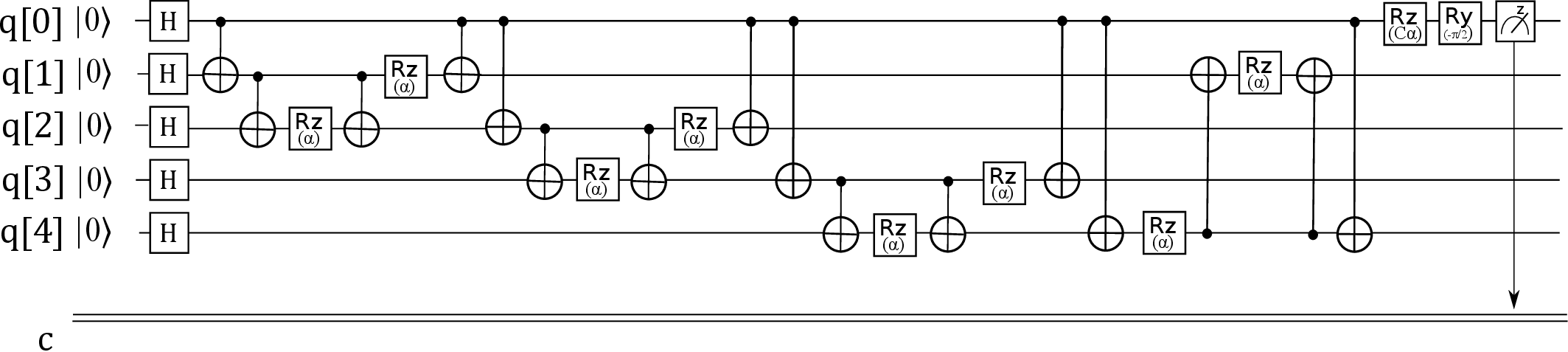}
\end{center}
\caption{Quantum protocol for studies of evolution of mean value of $\sigma^x_0$ in the case of Ising model on squared lattice in the magnetic field on a quantum computer. Here $C=9J$, $\alpha=2Jt/\hbar$. }
		\label{prot_square}
\end{figure}

In this subsection we present results of detection of the energy levels of the Ising model on squared lattice in the magnetic field on the quantum computer  ibmq-bogota. We study the following Hamiltonian
 \begin{eqnarray}
H=J\sigma^{z}_1\sigma^{z}_2+J\sigma^{z}_2\sigma^{z}_3+J\sigma^{z}_3\sigma^{z}_4+J\sigma^{z}_1\sigma^{z}_4+J\sigma^{z}_1+J\sigma^{z}_2+J\sigma^{z}_3+J\sigma^{z}_4.\label{hhh3388}
\end{eqnarray}
The expression for the total Hamiltonian reads
\begin{eqnarray}\label{Adelta90}
H_T=J\sigma^{z}_0(\sigma^{z}_1\sigma^{z}_2+\sigma^{z}_2\sigma^{z}_3+\sigma^{z}_3\sigma^{z}_4+\sigma^{z}_1\sigma^{z}_4+\sigma^{z}_1+\sigma^{z}_2+\sigma^{z}_3+\sigma^{z}_4+9).
\end{eqnarray}
The  constant is chosen to be $C=9J$. In this case the energy levels of $H$ are positive. On Fig. \ref{prot_square} we present quantum protocol for studies of evolution of  $\braket{\sigma^x_0}$ on quantum computer in the case of the total Hamiltonian (\ref{Adelta90}).

\begin{figure}
\begin{center}
\subcaptionbox{\label{ff2}}{\includegraphics[scale=0.75, angle=0.0, clip]{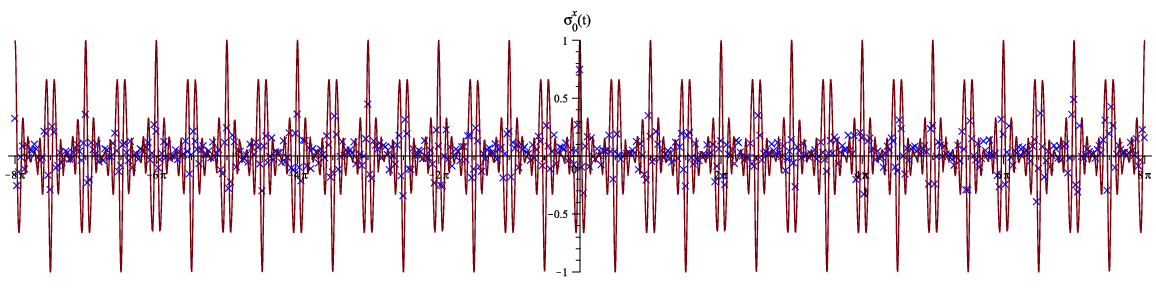}}
\subcaptionbox{\label{ff2}}{\includegraphics[scale=0.35, angle=0.0, clip]{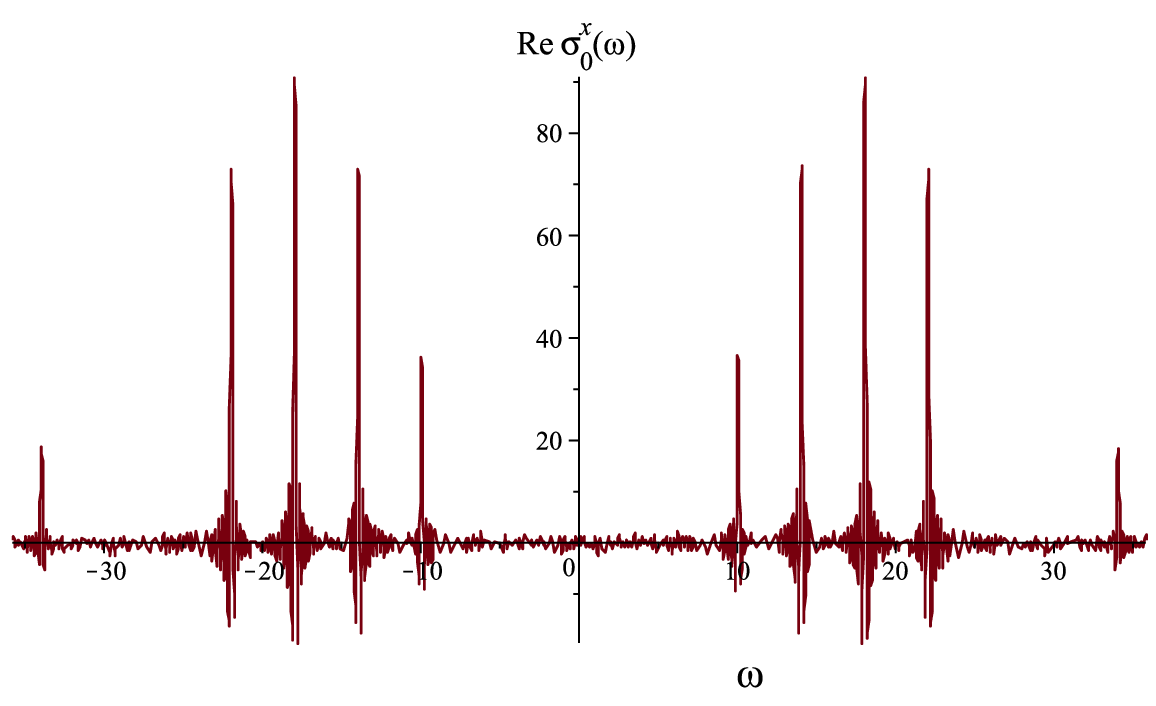}}
\subcaptionbox{\label{ff2}}{\includegraphics[scale=0.35, angle=0.0, clip]{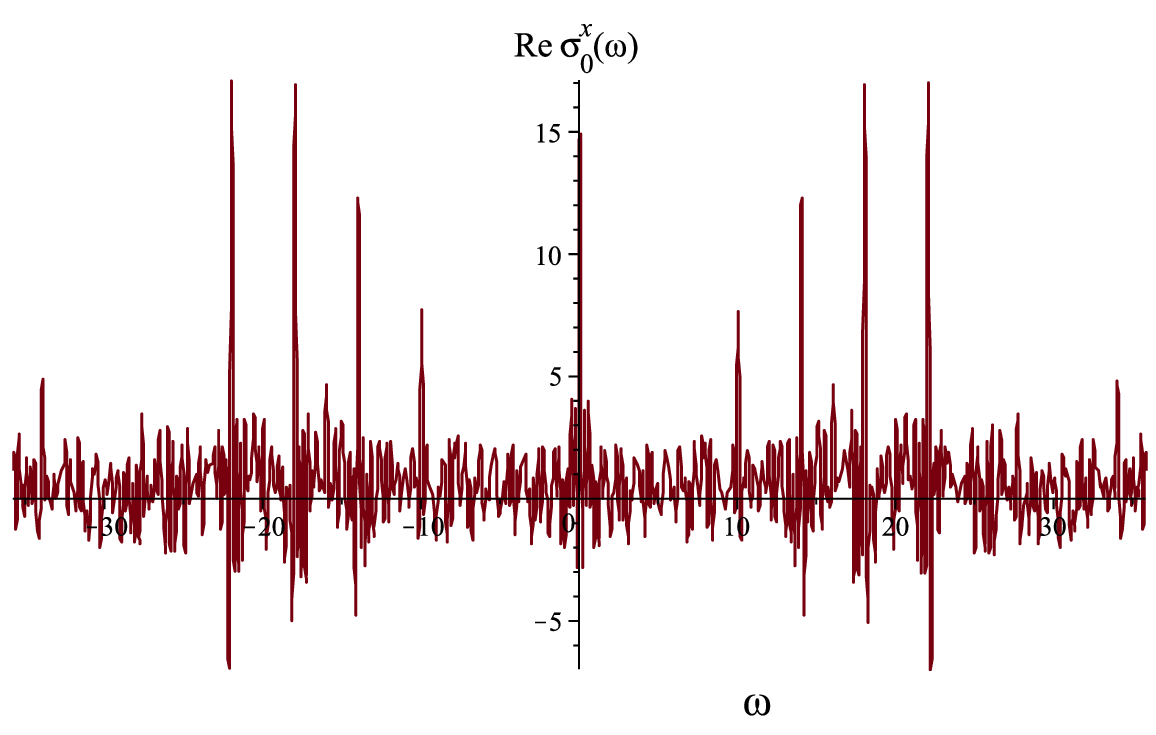}}
\end{center}
\caption{Results of quantifying  of the energy levels of Ising model on squared lattice  in the magnetic field  (\ref{hhh3388}). Evolution of the mean value $\braket{\sigma^x_0}$ detected on ibmq-bogota (a).  The real part of $\sigma^x_0(\omega)$ obtained on the basis of the results of calculations of $\sigma^x_0(t)$ on ibmq-qasm-simulator (b)  and  on ibmq-bogota (c).  The peaks of $\textrm{Re} \,\sigma^x_0(\omega)$ at  $\omega=\pm10$, $\omega=\pm 14$, $\omega=\pm 18$, $\omega=\pm 22$, $\omega=\pm 34$  correspond to energies $E_T=\pm5J$, $E_T=\pm7J$, $E_T=\pm9J$, $E_T=\pm11J$, $E_T=\pm17J$ of the total Hamiltonian (\ref{Adelta90}) and energies  $E=-4J$, $E=-2J$, $E=0$, $E=2J$, $E=8J$ of the Ising model on squared lattice  in the magnetic field (\ref{hhh3388}).}
		\label{fig99109}
\end{figure}

Quantum protocol Fig. \ref{prot_square} was implemented on ibmq-bogota for $\alpha/2=Jt/\hbar$ changing   from $-8\pi$ to $8\pi$ with the step $\pi/36$. The results of detection of the energy levels of the Ising model on squared lattice  in the magnetic field (\ref{hhh3388}) are presented on Fig. \ref{fig99109}. Similarly as in the previous examples for convenience we put  $J/\hbar=1$.
The  peaks of  $\textrm{Re} \,\sigma^x_0(\omega)$  at  $\omega=\pm10$, $\omega=\pm 14$, $\omega=\pm 18$, $\omega=\pm 22$, $\omega=\pm 34$  correspond to the total Hamiltonian $H_T$ energies $E_T=\pm5J$, $E_T=\pm7J$, $E_T=\pm9J$, $E_T=\pm11J$, $E_T=\pm17J$.  Similarly as in the previous cases the peak at $\omega=0$ Fig. \ref{fig99109} (c) is related with quantum errors and has not to be taken into account.
So, the detected energy levels of Ising square latice in the magnetic field  read $E=-4J$, $E=-2J$, $E=0$, $E=2J$, $E=8J$. The obtained results for the energy levels correspond the the analytical ones.

\section{Conclusions}

The method of detecting of the energy levels of a spin system on the basis of studies  of evolution of a probe spin has been proposed. The method provides possibility to estimate energy levels of many-spin system studying evolution of  the mean value of only one probe spin and  can be applied for arbitrary spin systems.

 We have realized the proposed algorithm on IBM's quantum computer ibmq-bogota and have detected  the energy levels of spin systems with Ising interaction (a spin chain in the magnetic field, triangle spin cluster,   Ising model on squared lattice  in the magnetic field). The results of quantum calculations (see Figs. \ref{fig99}, \ref{fig1099}, \ref{fig0099}, \ref{fig99109}) are in agreement with the theoretical ones.

On the basis of the obtained results  we conclude that the method is efficient even in the case of noisy quantum devices. The advantage of the proposed method is that for detection of the energy levels we use only one ancila qubit and do not need to measure the states of all qubits.   The quantum protocol Fig. \ref{fig:7} contains only measurement of one qubit which leads to reduction of the readout errors.
 The proposed method is efficient for estimation of the energy levels of many spin systems. Besides it is worth stressing that it is nontrivial combinatorial optimization problem to find minimal or maximal eigenvalue of the Ising model in the case when the constants  $J_{ij}$ in (\ref{Ising})  are different. Therefore the proposed method opens a possibility to achieve quantum supremacy in solving eigenvalue problem with development of multi-qubit quantum computers.

\section*{Acknowledgments}
This work was supported by Project 2020.02/0196 (No. 0120U104801) from National Research Foundation of Ukraine.

\end{document}